\documentclass[twocolumn,aps]{revtex4}

\usepackage{graphicx,graphics}
\usepackage{amsmath}
\usepackage{amsfonts}
\usepackage{amssymb}
\usepackage{amsthm}

\begin{document}

\title{Numerical Study of the Correspondence Between the Dissipative and Fixed Energy Abelian Sandpile Models}

\author{Su.S. Poghosyan$^1$, V.S. Poghosyan$^2$, V.B. Priezzhev$^3$ and P. Ruelle$^2$}
\affiliation{
$^1$Institute for Informatics and Automation Problems NAS of Armenia, 375044 Yerevan, Armenia\\
$^2$Institut de Physique Th\'{e}orique, Universit\'{e} catholique de Louvain, B-1348 Louvain-La-Neuve, Belgium\\
$^3$ Laboratory of Theoretical Physics, Joint Institute for Nuclear Research, 141980 Dubna, Russia
}

\begin{abstract}
{
We consider the Abelian sandpile model (ASM) on the large square lattice with a single dissipative site (sink).
Particles are added by one per unit time at random sites and the resulting density of particles is calculated as a function of time.
We observe different scenarios of evolution depending on the value of initial uniform density (height) $h_0=0,1,2,3$.
During the first stage of the evolution,  the density of particles increases linearly.
Reaching a critical density $\rho_c(h_0)$, the system changes its behavior sharply and relaxes exponentially to the stationary state of the ASM
with $\rho_s=25/8$. We found numerically that $\rho_c(0)=\rho_s$ and $\rho_c(h_0>0) \neq \rho_s$.
Our observations suggest that the equality $\rho_c=\rho_s$ holds  for more general initial conditions with non-positive heights.
In parallel with the ASM, we consider the conservative fixed-energy Abelian sandpile model (FES).
The extensive Monte-Carlo simulations for $h_0=0,1,2,3$ have confirmed
that in the limit of large lattices $\rho_c(h_0)$ coincides with the threshold density $\rho_{th}(h_0)$ of FES.
Therefore, $\rho_{th}(h_0)$ can be identified with $\rho_s$ if the FES starts its evolution with non-positive uniform height $h_0 \leq 0$.}
\end{abstract}

\maketitle


\section{Introduction}

A long-standing discussion of the comparative critical properties of the dissipative Abelian sandpile model (ASM) \cite{btw}
and the conservative fixed-energy sandpile (FES) \cite{VDMZ} has gained recently renewed impetus due to the works by Fey et al \cite{Fey}.
The point of discussion in a laconic form can be reduced to the single question: given the same lattice
with open and closed boundary conditions,
whether the stationary density of the dissipative ASM  $\rho_s$ coincides with the threshold density of the FES $\rho_{th}$ ?
Using large-scale simulations on the square lattice, the authors of works \cite{Fey, Fey2} gave a negative answer to this
question and supported their numerical findings by exact solutions for some graphs of higher symmetry.
A more detailed answer contains a description of the average density of grains $\rho(\tau)$ for the given density of added particles $\tau$
of the dissipative sandpile:
\begin{equation}
\rho(\tau) = \frac{\text{\#grains at time } t}{N^2}, \quad \tau=\frac{t}{N^2}.
\end{equation}
where $N$ is the linear size of the lattice.
It was shown in \cite{Fey} that $\rho(\tau)$
for the ASM exhibits a transition at the point $\rho_c$ which coincides with the threshold density of the FES:  $\tau =\rho_c = \rho_{th}$.
For $\tau > \rho_{c}$, the system continues evolution and its density tends to the stationary value $\rho_s$ of the ASM when $\tau \rightarrow \infty$.

In this article, we continue the analysis of the correspondence between critical points of the ASM and FES.
Confirming the main result of \cite{Fey} on the transition point at $\rho_c = \rho_{th}$,
we present arguments against the general claim $\rho_{th} \neq \rho_{s}$ for the two-dimensional square lattice.
The focus of the work will be the dependence of the transition point on the initial conditions.
Specifically, we study the nearly closed sandpile, namely the ASM on the square lattice with periodic boundary conditions and a single dissipative site.
One of the results suggested by this study is that there exist initial conditions which lead during
the evolution to the transition point $\rho_c$ coinciding with $\rho_s=25/8$, and the subsequent evolution does not change this density.
On the other hand, our simulations of the  ASM and FES show that the transition point of the nearly closed ASM coincides
for large lattices with the threshold density $\rho_{th}$ of the FES for all considered initial conditions.
To fix our notations, we will specify the two-dimensional ASM as in \cite{ddhar}, i.e. all stable sites have heights $h_i \leq 4$ whereas
sites with $h_i \geq 5$ are unstable and topple.

Traditionally, empty sites $(h_i=0)$ and negative heights are not considered neither analytically nor numerically because they exist
in transient states only and vanish when a system approaches criticality. Nevertheless, these sites may affect the evolution of the ASM from an
initial state to the transition point.
The dynamics of the nearly closed ASM is very close to that of the FES as dissipation through the single site is strongly restricted.
Non-ergodicity of the FES causing a dependence of the threshold density upon the initial conditions has been discussed recently in \cite{Park}.
A conclusion from this discussion was: ``the self-organized criticality exhibited by the ASM has nothing to do with phase transitions in the FES'' \cite{Park}.
Being reasonable, this statement does not exclude however specific initial conditions whose evolution leads the FES exactly to the stationary density of the
ASM $\rho_s=25/8$.

In what follows, we analyze five initial conditions with uniform heights $h_0=0,1,2,3$ for all lattice sites beside the sink.
The density $\rho$ as a function of the density of added particles $\tau$ behaves differently for different $h_0$ and
has a characteristic kink at $\rho_c(h_0)$, also depending on $h_0$. For $h_0=1$, the value $\rho_c(1) = 3.12528\ldots$
is close to the threshold density of the FES obtained by Fey et al \cite{Fey}.
For $h_0=0$, the density $\rho$ increases linearly with $\tau$, reaches the value $\rho_s \simeq 25/8$ and remains constant later.
In Section 2, we give the definition of the model and remind the origin of the conjectural exact value $\rho_s = 25/8$.
In Section 3 we present details of the numerical procedure and describe five different scenarios of evolution for the initial conditions $h_0=0,1,2,3$.
The special case $h_0=0$ is considered in more details.

In parallel with the ASM, we present  simulations of the FES for finite $N \times N$ square lattice with various $N$
and demonstrate the critical behavior of the threshold density $\rho_{th}(h_0)$.
Namely, we find numerically the finite size correction $1/N^{\gamma}$ with critical index $\gamma=\gamma(h_0)$ that depends on the initial conditions.

\section{The model and critical density}

We consider the standard Abelian sandpile model \cite{btw} on the $N\times N$ square lattice.
A peculiarity of our consideration is the nearly closed boundary conditions: the lattice is a torus with a single selected site $i_0$.
The height $z_i$ at any site beside  $i_0$ takes values $0,1,2,3,4$ in stable configurations.
Particles are added by one  at a random site increasing the height at that site by one.
If the height exceeds $4$, then the site becomes unstable and topples, transferring one particle to each of four neighboring sites.
The site  $i_0$ serves as a sink where particles disappear. During a long evolution, the system gets eventually into a set of recurrent configurations.
The theory of recurrent states of the ASM was developed by Dhar \cite{ddhar}.
One of important consequences of this theory is the existence of a mapping of the set of recurrent configurations
onto the set of spanning trees of the same lattice.
The mapping allows one to find the height probabilities in the recurrent state and therefore to find the stationary density $\rho_s$.
Majumdar and Dhar \cite{Dhar} calculated the probability of height $h=1$ in the thermodynamic limit
\begin{equation}
P_1=\frac{2}{\pi^2}-\frac{4}{\pi^3}.
\label{P1}
\end{equation}
The probabilities $P_2$, $P_3$ and $P_4=1-P_1-P_2-P_3$ were found in \cite{Priez}:
\begin{eqnarray}
&&P_2=\frac{1}{2}-\frac{3}{2\pi}-\frac{2}{\pi^2}+\frac{12}{\pi^3}+\frac{I_1}{4}, \label{PP2} \\
&&P_3=\frac{1}{4}+\frac{3}{2\pi}+\frac{1}{\pi^2}-\frac{12}{\pi^3}-\frac{I_1}{2}-\frac{3I_2}{32}, \label{PP3} \\
&&P_4=\frac{1}{4}-\frac{1}{\pi^2}+\frac{4}{\pi^3}+\frac{I_1}{4}+\frac{3I_2}{32}. \label{PP4}
\end{eqnarray}
Here $I_{\nu}$, $\nu=1,2$ are integrals
\begin{equation}
I_{\nu}=\frac{1}{(2\pi)^4}\int\!\!\!\!\!\int\!\!\!\!\!\int\!\!\!\!\!\int_{0}^{2\pi}\!\!\!\!\!
\frac{i \sin \beta_1 \det M_{\nu} \, d\alpha_1 d\alpha_2 d\beta_1 d\beta_2}
{D(\alpha_1,\beta_1)D(\alpha_2,\beta_2)D(\alpha_1+\alpha_2,\beta_1+\beta_2)},
\label{I}
\end{equation}
where
\begin{equation}
D(\alpha,\beta)=2-\cos(\alpha)-\cos(\beta)
\label{D}
\end{equation}
and $M_1, M_2$ are the matrices
\begin{equation}
M_1=
\begin{pmatrix}
1 & 1 & e^{i\alpha_2} & 1 \\
3 & e^{i(\beta_1+\beta_2)} & e^{i(\alpha_2-\beta_2)} & e^{-i\beta_1} \\
4/\pi-1 & e^{i(\alpha_1+\alpha_2)} & 1 & e^{-i\alpha_1} \\
4/\pi-1 & e^{-i(\alpha_1+\alpha_2)} & e^{2i\alpha_2} & e^{i\alpha_1}
\end{pmatrix},
\label{M1}
\end{equation}
\begin{equation}
M_2=
\begin{pmatrix}
e^{i\beta_2} & e^{-i(\alpha_1+\alpha_2)-i(\beta_1+\beta_2)} & e^{i\beta_1} \\
e^{-i\alpha_2} & 1 & e^{-i\alpha_1} \\
e^{i\alpha_2} & e^{-2i(\alpha_1+\alpha_2)} & e^{i\alpha_1}
\end{pmatrix}.
\label{M2}
\end{equation}

Later on, an exact relation between $P_2$ and $P_3$ was proved in \cite{jpr}, thereby eliminating one of the two integrals.
A conjecture on the value of the remaining one, based on its numerical evaluation to twelve decimal places,
then led to the following values for the $P_i$ \cite{jpr},
\begin{eqnarray}
&& P_2 = \frac{1}{4} - \frac{1}{2\pi} - \frac{3}{\pi^2} + \frac{12}{\pi^3},\\
&& P_3 = \frac{3}{8} + \frac{1}{\pi} - \frac{12}{\pi^3},
\end{eqnarray}
and for the stationary density,
\begin{equation}
\rho_s = P_1 + 2 P_2 + 3 P_3 + 4 P_4 = \frac{25}{8}.
\label{crit}
\end{equation}
This value of $\rho_s$ was first suggested by Grassberger \cite{Grassberger} based on a high precision calculation of the integrals $I_1,I_2$.
The present numerical accuracy reaches $10^{-25}$, and forms the basis for the conjecture that $\rho_s = 25/8$ exactly.

Another conjecture, coined in \cite{Fey} as the "density conjecture", identifies $\rho_s$ with the threshold density of the FES, $\rho_{th}$ \cite{VDMZ}.
The threshold density is defined as the average maximal number of randomly dropped particles per site which allows the relaxation of the closed sandpile to stop.
The value $\rho_{th}$ marks the border of stability:
for any $\rho > \rho_{th}$, an avalanche process started by adding a particle never stops with probability one.
The conjecture $\rho_{th}=\rho_{s}$ was supported  by numerical calculations \cite{VDMZ}.
However, recently, Fey et al \cite{Fey} have obtained $\rho_{th}=\rho_{s}+\Delta$, where $\Delta=0.000288$ to six decimal places.
Together with the claim by Park \cite{Park} quoted in Introduction, this makes the correspondence between critical points of the ASM and FES rather problematic.

We consider here a nearly closed sandpile with a single open site $i_0$.
An idea is to retrace carefully the process of adding particles from an initial state to the crirical state of the ASM.
At the initial stage of the process, the probability of dissipation via the sink at $i_0$ is negligibly small for the large lattices
and the evolution of the system almost coincides  with the dynamics of the FES.
When the density approaches a critical value, the large avalanches appear and the probability of dissipation increases.
If the transition to the dissipative regime is sufficiently sharp, we can identify the transition point with  $\rho_{c}$.
According to the process described in \cite{Fey}, the system evolves with further adding of particles and reaches eventually
the stationary density of the ASM $\rho_{s}$.
The main difference from previous studies is in the initial conditions.
Beside the usual initial conditions $1 \leq h_0 \leq 3$, we consider the empty lattice $h_0=0$ and show that just this condition
leads to the transition point $\rho_c = \rho_{s}$.

\section{Simulations for ASM with one sink}

Consider the Abelian sandpile model with one sink on a $N \times N$ square lattice $\mathcal{L}$ with periodic boundary condition in both directions.
The initial configuration is uniform, $h_i = h_0$ with $h_0=0,1,2,3$.
We add particles by one per unit time at a random site and the resulting density of particles is measured as a function of time.
At the first stage of the evolution, the system loses negligibly small fraction of particles to the sink,
and the density of particles increases with time linearly.
Eventually, the system reaches some critical density $\rho_c(h_0)$ where it  starts losing a macroscopic amount of particles.
At this point, the density profile changes sharply and the density tends monotonically to the stationary density of the ASM $\rho_s=25/8$.
At the transition point, the density $\rho(\tau)$ has a kink which allows one to determine $\rho_c(h_0)$ with reasonable accuracy.

\subsection{$h_0=3$}

During the initial stage of evolution, the number of particles in the system grows linearly with time $\tau$.
Rare dissipative events occur only if the cluster of sites with $h=4$ is situated in a close vicinity of $i_0$ and the added particle hits there.
When the density of particles reaches a threshold level, a kind of percolation appears which can be termed a "weak percolation".
It implies that sites with $h=4$ do not percolate yet, but additional sites increasing their height during an avalanche,
produce together with them a percolation set. The point of the weak percolation is rather low ($\rho_c \approx 3.09$ in our simulations).
Therefore, the density of particles continues a slow increase despite the big avalanches spreading through the system.
The weak percolation point is visible in Fig.2 as a kink following the linear part of the function $\rho(\tau)$.

\begin{figure}[!h]
\includegraphics[width=80mm]{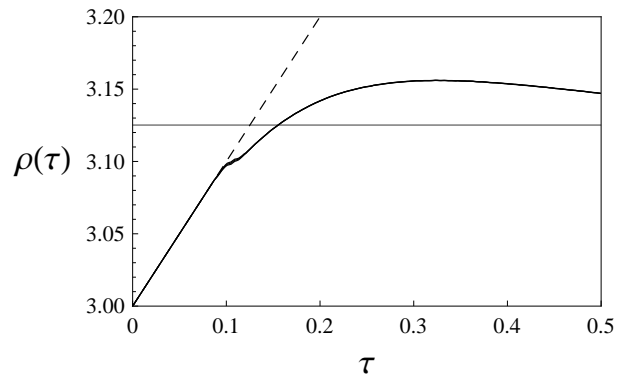}
\caption{The evolution of the dissipative ASM with one sink for $h_0=3$, number of samples -- 1000, $N=500$.}\label{fig3}
\end{figure}

\subsection{$h_0=2$}

Due to lowering of the initial density, the percolation picture of the sites with $h=4$ becomes more pronounced, getting close to the usual site percolation.
The percolation dynamics versus the avalanche dynamics for different background densities in the FES has been discussed recently by Park \cite{Park}.
It was demonstrated that the critical density of transition into the unstable state strongly depends on the background density and may differ considerably
both from accepted values of $\rho_s$  and  $\rho_{th}$. Our results for $h_0=2$ confirm this conclusion. The density of particles $\rho(\tau)$
grows strictly linearly in time up to the transition point $\rho_c \approx 3.134$ which can be associated, due to rare dissipation events,
with the critical density of the FES for the same initial condition.
At the transition point, the avalanche mechanism is activated and the dissipation reduces density towards $\rho_s$.
It may seem strange that lowering of the initial density from $h_0=3$ to $h_0=2$ leads to the increase of the transition point from
$\rho_c \approx 3.09$ to $\rho_c \approx 3.134$.
However, this is the result of the competition between larger number of percolation sites with $h=4$ for $h_0=2$ and the reduced density of particles
outside the percolation cluster. Of course, the value of the transition point cannot be derived directly from the site-percolation critical probability
$P_s=0.592...$, because the distribution of heights outside the percolation cluster remains unknown.

\begin{figure}[!t]
\includegraphics[width=80mm]{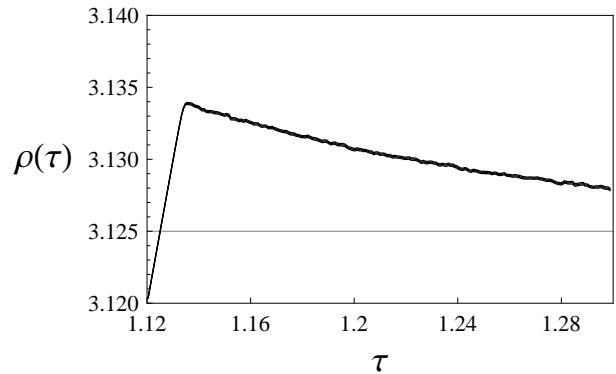}
\caption{The evolution of the dissipative ASM with one sink for $h_0=2$, number of samples -- 1000, $N=500$.}\label{fig2}
\end{figure}

\subsection{$h_0=1$}

\begin{figure}[!b]
\includegraphics[width=80mm]{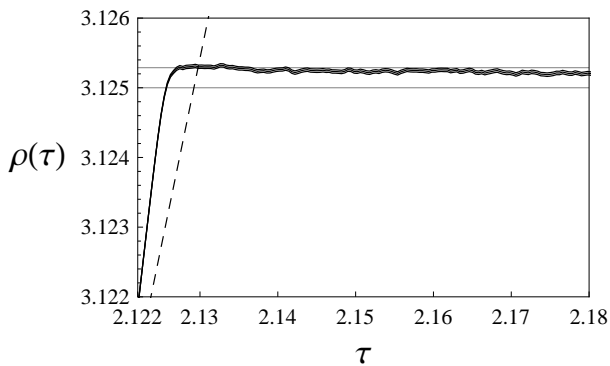}
\caption{The evolution of the dissipative ASM with one sink for $h_0=1$, number of samples -- 1000, $N=500$.
Two horizontal lines correspond to values $\rho_{th}(0)=3.125$ and $\rho_{th}(1)=3.125288\ldots$\,.
}\label{fig1}
\end{figure}

This case, basically, is very similar to the previous one.
However, a tendency of increasing of the transition point with lowering of the initial density observed in the case $h_0=2$ is reversed.
Remarkably, the value $\rho_c \approx 3.12528$ for the relatively small lattice $N=500$ is very close (after reduction by one) to the threshold
value of the FES $\rho_{th}=2.125278 \pm 0.0000004$ reported in \cite{Fey} for the lattice $N=512$.
This supports the idea that the nearly closed sandpile behaves almost identically to the FES during the non-dissipative part of evolution.
Indeed, the simulations in \cite{Fey} were started with the initial conditions
$\zeta=0$ for the sandpile model with stable occupation numbers $\zeta = 0,1,2,3$ which correspond to heights $h=1,2,3,4$ in our notations.
Therefore, $\zeta=0$ for the FES in \cite{Fey} corresponds to $h_0=1$ in our simulations. Analogously, our $h_0=0$ corresponds to $\zeta=-1$.

\subsection{$h_0=0$}

\begin{figure}[!b]
\includegraphics[width=80mm]{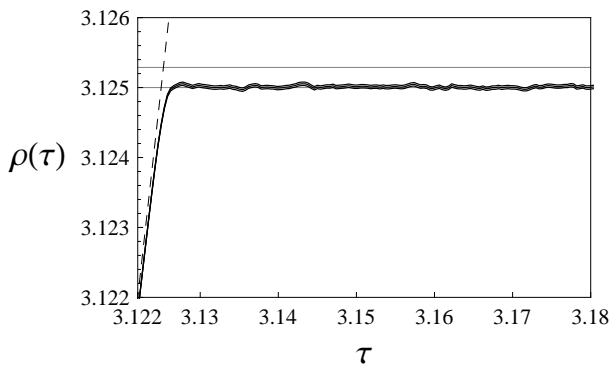}
\caption{The evolution of the dissipative ASM with one sink for $h_0=0$, number of samples -- 1000, $N=500$.
Two horizontal lines correspond to values $\rho_{th}(0)=3.125$ and $\rho_{th}(1)=3.125288\ldots$\,.
}\label{fig0}
\end{figure}

The initial condition $h_0=0$ is special and turns out to be different from the other initial conditions.
The similarity between dynamics of the FES and the nearly closed ASM observed in the previous cases allows us to use an important
statement proved for the FES: if a configuration stabilizes after an avalanche process, there is some site that has not toppled \cite{Tar88}
(see a similar theorem in \cite{FdBMR09,FLP10}). We may expect (and we do observe this in our simulations) that there is a set of sites
with non-zero density which never topple during the evolution of the nearly closed ASM from the initial state up to the threshold value where
all sites topple ultimately. Then, the occupation number at these sites does not affect the dynamics, giving at the the same time a contribution
to the total density. If so, the threshold point $\rho_c \approx 3.1252$ found for $h_0=1$ can be lowered for $h_0=0$ to the critical point of the ASM.
Our simulations confirm this expectation. Adding particles to the empty lattice, we observe as above the linear growth of density $\rho(\tau)$
which indicates almost non-dissipative character of evolution. The linear part changes sharply at the threshold value $\rho_c \approx 3.125$
into a function slightly fluctuating around $\rho(\tau)=const$. The coincidence of $\rho_c$ and $\rho_s$ implies that the percolation
dynamics does not dominate anymore and the avalanche process approaches its standard critical behavior.

Among the initial conditions that we have considered, $h_0=0$ is the only one for which the threshold density $\rho_s$
appears to be close or equal to the stationary density $25/8$.
We expect that this remains true for other initial conditions with non-positive heights.

\section{The threshold density of FES}

Along with the nearly closed sandpile, we considered the FES for initial conditions $h_0=0,1,2,3$ to confirm the relation
$\rho_c=\rho_{th}$ and to show the power-law convergence of the threshold density to its asymptotical value.
To do that, we have considered $N \times N$ lattices for various $N$, and periodic boundary conditions in both directions.
On each of these lattices, we run $10^7$ samples of fixed energy sandpile model and measure average threshold density $\rho_{th}(N, h_0)$
and its standard deviation.
After the extrapolation of $\rho_{th}(N, h_0)$ we found numerically the asymptotic value of the threshold density and its finite-size corrections for large $N$.
The accuracy $\delta \rho_{th}(N,h_0)$ of the result $\rho_{th}(N,h_0)$ is estimated via its standard deviation $D\rho_{th}(N,h_0)$ by the following formula:
\begin{equation}
\delta\rho_{th}(N, h_0) \sim \frac{D\rho_{th}(N)}{\sqrt{\text{\#samples}}}
\end{equation}

\begin{figure}[!h]
\includegraphics[width=80mm]{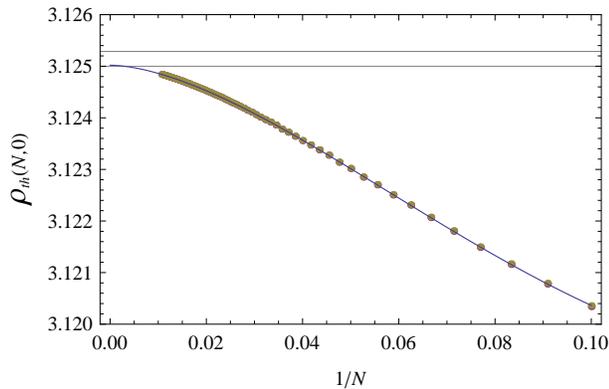}
\caption{The threshold densities of the fixed energy sandpile for $h_0=0$, number of samples -- $10^7$ and $N=10,11,\ldots,100$.
Two horizontal lines correspond to values $\rho_{th}(0)=3.125$ and $\rho_{th}(1)=3.125288\ldots$\,.
}\label{figFES0}
\end{figure}

\begin{figure}[!h]
\includegraphics[width=80mm]{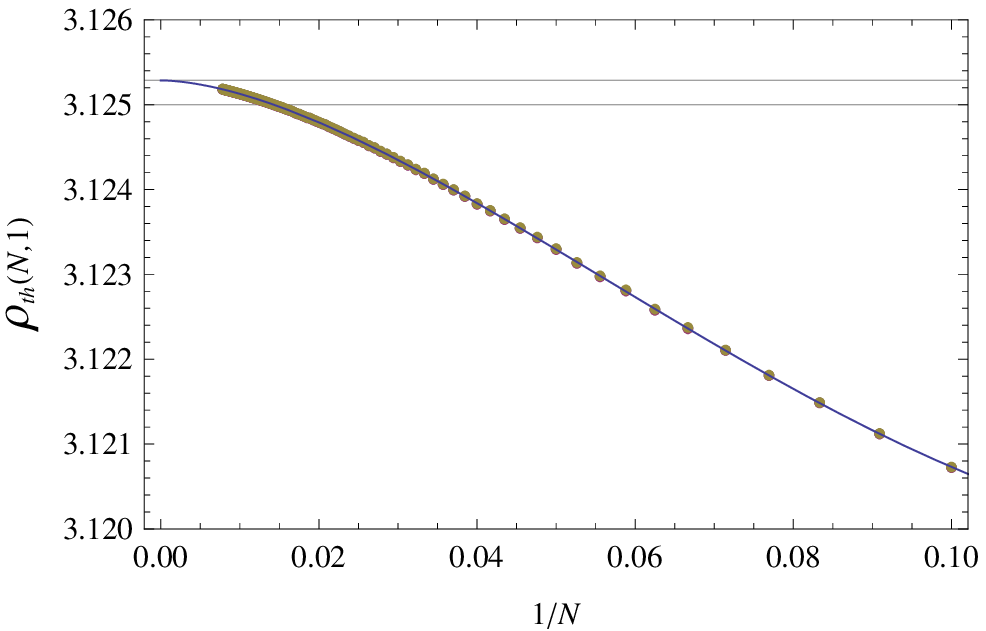}
\caption{The threshold densities of the fixed energy sandpile for $h_0=1$, number of samples -- $10^7$ and $N=10,11,\ldots,100$.
Two horizontal lines correspond to values $\rho_{th}(0)=3.125$ and $\rho_{th}(1)=3.125288\ldots$\,.
}\label{figFES1}
\end{figure}

\begin{figure}[!h]
\includegraphics[width=80mm]{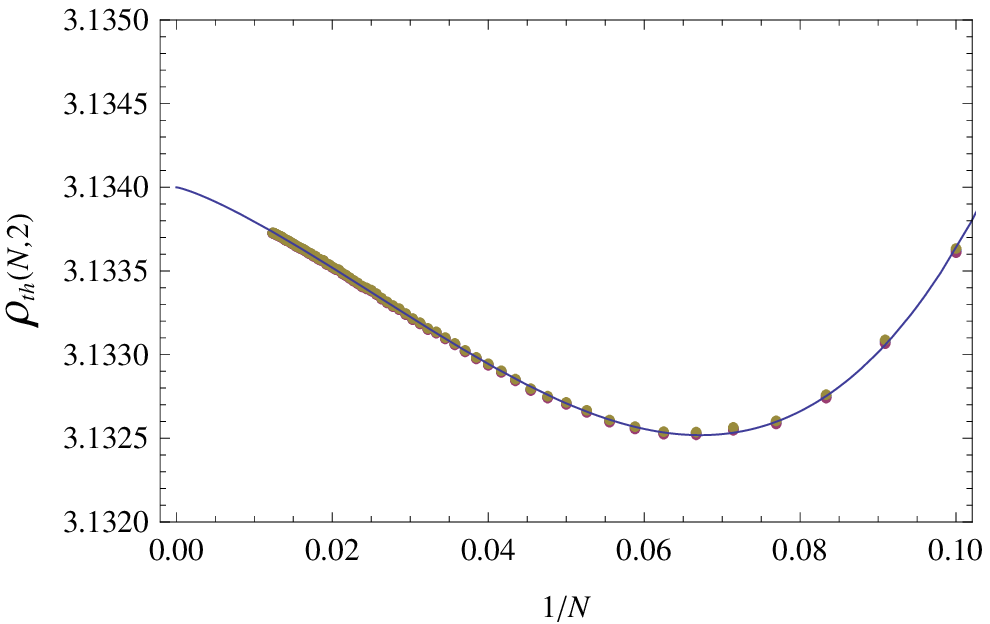}
\caption{The threshold densities of the fixed energy sandpile for $h_0=2$, number of samples -- $10^7$ and $N=10,11,\ldots,100$.}
\label{figFES2}
\end{figure}

\begin{figure}[!h]
\includegraphics[width=80mm]{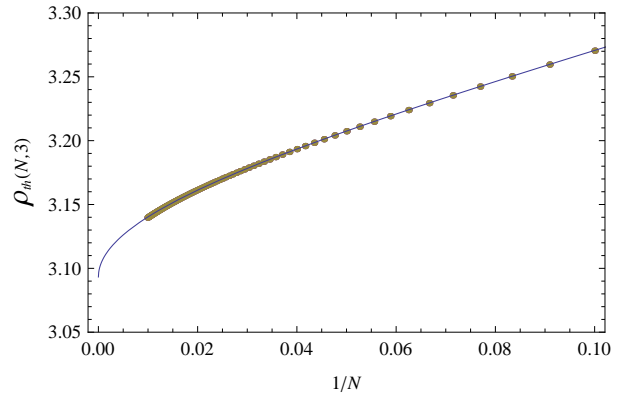}
\caption{The threshold densities of the fixed energy sandpile for $h_0=3$, number of samples -- $10^7$ and $N=10,11,\ldots,100$.}
\label{figFES3}
\end{figure}

The results of the simulations on FES with various initial conditions are shown in Figs. \ref{figFES0}--\ref{figFES3}.
The sizes of dots in the figures are proportional to their accuracy of the simulations.
For $h_0=0$ we found
\begin{equation}
\rho_{th}(N, 0) \simeq 3.1250(1) - 0.5(5) N^{-\frac{7}{4}} + 1.6(1) N^{-\frac{5}{2}}.
\end{equation}
For $h_0=1$ we found
\begin{equation}
\rho_{th}(N, 1) \simeq 3.12528(5) - 0.5(5) N^{-\frac{7}{4}} + 1.6(1) N^{-\frac{5}{2}}.
\end{equation}
For $h_0=2$ we found
\begin{equation}
\rho_{th}(N, 2) \simeq 3.13400(2) - 0.06(5) N^{-\frac{5}{4}} + 18.5(4) N^{-\frac{15}{4}}.
\end{equation}
For $h_0=3$ we found
\begin{equation}
\rho_{th}(N, 3) \simeq 3.0938(7) + 0.457(4) N^{-\frac{1}{2}} + 1.010(7) N^{-\frac{3}{2}}.
\end{equation}

\section{Conclusion}

The results of our numerical simulations support the following three statements:
(1) the density of sand held by the lattice with a single dissipative site (or with dissipation rate tending to zero),
as a function of the quantity of sand deposed, shows a discontinuity at a value $\rho_c$,
which can be identified with the threshold density of the corresponding FES (with the same initial condition);
(2) the critical value $\rho_c$ depends on the initial condition, and consequently, there is presumably also a range of values for the FES threshold density,
also depending on the initial condition and potentially on the way sand is dropped; (3) for the specific uniform initial configuration $h_0=0$,
and probably for others as well, the critical is close or equal to the stationary density, $\rho_c=\rho_s=25/8$.
One can expect that if the system starts its evolution from sufficiently deep initial conditions,
i.e. from initial condition with large negative heights, the stationary value $25/8$ coincides with $\rho_c$ and  $\rho_{th}$ exactly.

\section*{ACKNOWLEDGMENTS}
This work was supported by the Russian RFBR grant No 09-01-00271a, and by the Belgian Interuniversity Attraction Poles Program P6/02,
through the network NOSY (Nonlinear systems, stochastic processes and statistical mechanics).
P.R. is Senior Research Associate of the Belgian National Fund for Scientific Research (FNRS).
The Monte-Carlo simulations were performed on Armenian Cluster for High Performance Computation (ArmCluster, www.cluster.am).


\begin{thebibliography}{99}

\bibitem{btw} P. Bak, C. Tang and K. Wiesenfeld, Phys. Rev. Lett. {\bf 59}, 381 (1987).

\bibitem{VDMZ} A. Vespignani, R. Dickman, M.A. Mu\~{n}noz, and S. Zapperi, Phys. Rev. E {\bf 62}, 4564-4582 (2000).

\bibitem{Fey} A. Fey, L. Levine and D.B. Wilson, Phys. Rev. Lett. {\bf 104}, 145703 (2010).

\bibitem{Fey2} A. Fey, R. Meester and F. Redig, Ann. Prob. {\bf 37}, 654-675 (2009).

\bibitem{jpr} G. Piroux and P. Ruelle, Phys. Lett. {\bf B 607}, 188 (2005);
M. Jeng, G. Piroux and P. Ruelle, J. Stat. Mech. P10015 (2006).

\bibitem{ipd} E.V. Ivashkevich and V.B. Priezzhev, Physica {\bf A 254}, 97-116 (1998);
D. Dhar, Physica {\bf A 369}, 29 (2006).

\bibitem{ddhar} D. Dhar, Phys. Rev. Lett. {\bf 64}, 1613 (1990).

\bibitem{Dhar} S.N. Majumdar and D. Dhar, J. Phys. A: Math. Gen. {\bf 24}, L357-L362 (1991).

\bibitem{Priez} V.B.Priezzhev, J. Stat. Phys {\bf 74}, 955 (1994).

\bibitem{correlations} V.S. Poghosyan, S.Y. Grigorev, V.B. Priezzhev and P. Ruelle, Phys. Lett. B {\bf 659}, 768 (2008);\\ J. Stat. Mech. P07025 (2010).

\bibitem{Grassberger} P. Grassberger, private comunications.

\bibitem{Park} Su-Chan Park, arXiv:1001.3359v1 [cond-mat.stat-mech] 19 Jan 2010.

\bibitem{Tar88} G. Tardos, SIAM J. Discrete Math. {\bf 1}, 397–398 (1988).

\bibitem{FdBMR09} A. Fey, R. Meester and F. Redig, Ann. Prob. {\bf 37}, 2, 654-675 (2009).

\bibitem{FLP10} A. Fey, L. Levine and Y. Peres, J. Stat. Phys. {\bf 138} 143-159 (2010).

\end{thebibliography}
\end{document}